\documentclass[10pt,english]{article}
\usepackage{times}
\usepackage[T1]{fontenc}
\usepackage[latin1]{inputenc}
\usepackage{amsmath}
\usepackage{amssymb}

\makeatletter

\newcommand{\noun}[1]{\textsc{#1}}

\usepackage{slashed}

\usepackage{babel}
\makeatother
\begin{document}

\title{\textbf{The exact eigenstates of the neutrino mass matrix without
CP-phase violation}}

\author{\noun{ADRIAN} PALCU}

\date{\emph{Department of Theoretical and Computational Physics - West
University of Timi\c{s}oara, V. P\^{a}rvan Ave. 4, RO - 300223 Romania}}

\maketitle
\begin{abstract}
In this paper we obtain the exact mass-eigenstates of the Majorana
physical neutrinos. We start by taking into account a general $3\times3$
mass matrix without any CP-phase violation. It is then diagonalized
by exactly solving an appropriate set of equations. The solution supplies
straightforwardly the mass eigenvalues depending on the diagonal entries
and mixing angles. Finally, the consequences of these analytical expressions
are discussed assuming various phenomenological restrictions such
as conserving the global lepton number $L=L_{e}-L_{\mu}-L_{\tau}$
and the $\mu-\tau$ interchange symmetry. The minimal absolute mass
in the neutrino sector is also obtained since the two plausible scenarios
invoked above are employed.


\end{abstract}

\section{Introduction}

Detecting neutrino oscillations \cite{key-1} in atmospheric and solar
neutrino fluxes (and then observing the same phenomenon in reactor
and accelerator experiments) stands as a milestone in particle physics
of the last decade. This compelling experimental evidence proves that
a suitable theory must rely - among many other ingredients - on the
fact that lepton flavor can not be conserved. Therefore, one must
take into account that neutrinos mix (as quarks do). In other words
they can oscillate into one another. That is: when the neutrino flavor
is subject of a measurement in a neutrino flux, one obtaines different
results if a macroscopic distance separates the detectors interacting
with that flux. On the other hand, the Quantum Field Theory suggests
that the neutrinos have to be represented by fermion massive fields,
let them be of Majorana or Dirac type. At the same time, the unified
theory design to describe the interactions must explain why neutrino
masses are so tiny when compared to the charged lepton ones. A striking
feature also arises since the data favor surprisingly large atmospheric
and solar mixing angles, in contrast with the quark mixing pattern.

All these features make difficult the attempts to precisely establish
the structure of the neutrino mass matrix able to fit the available
data supplied by global analysis \cite{key-2}. Specific textures
have been proposed taking into consideration certain discrete symmetries
that could govern the lepton families. Some of these symmetries are
compatible with the elegant seesaw mechanism \cite{key-3} designed
to predict the observed small order of magnitude for the masses of
physical neutrinos. 

A different approach consists of advancing certain gauge models that
give rise to specific Yukawa sectors able to supply a concrete mass
matrix. Encouraged by the success of such a strategy in the case of
a particular 3-3-1 gauge model, we propose here an analytical diagonalization
of a general neutrino mass matrix just by taking into account arbitrary
diagonal entries instead of the particular ones considered in the
3-3-1 model previous papers of the author \cite{key-4} based on the
general method of exactly solving gauge models with high symmetries
\cite{key-5}.

The paper is organized as follows. In Section 2 the theoretical framework
\cite{key-6} of the neutrino mixing is briefly presented with the
standard notations of the field. Section 3 deals with the exact neutrino
mass eigenstates and eigenvalues obtained by solving a set of equations
corresponding to the diagonalization of the general mass matrix. Certain
phenomenological restrictions are introduced in Section 4 where the
main results of the paper are presented. Last section is devoted to
conclusions and comments on the obtained results.

\section{The neutrino mass matrix}

We start by assuming the neutrino mixing formula: $\nu_{\alpha L}(x)=\sum_{i=1}^{3}U_{\alpha i}\nu_{iL}(x)$,
where $\alpha=e,\mu,\nu$ label the flavor space (flavor gauge eigenstates),
while $i=1,2,3$ denote the massive physical eigenstates. We consider
throughout this paper the physical neutrinos as Majorana fields, \emph{i.e.}
$\nu_{iL}^{c}(x)=\nu_{iL}(x)$. The mass term in the Yukawa sector
of any gauge unified theory that generate Majorana neutrinos stands: 

\begin{equation}
\mathcal{-L}_{Y}=\frac{1}{2}\bar{\nu}_{L}M\nu_{L}^{c}+H.c\label{Eq. 1}\end{equation}
with $\nu_{L}=\left(\begin{array}{ccc}
\nu_{e} & \nu_{\mu} & \nu_{\tau}\end{array}\right)_{L}^{T}$ where the superscript $T$ denotes ''transposed'' . The complex
mixing matrix $U$ that diagonalizes the mass matrix $M$ in the manner
$U^{+}MU=m_{ij}\delta_{j}$ has in the standard parametrization the
form: 

\begin{equation}
U=\left(\begin{array}{ccc}
c_{2}c_{3} & s_{2}c_{3} & s_{3}e^{-i\delta}\\
-s_{2}c_{1}-c_{2}s_{1}s_{3}e^{i\delta} & c_{1}c_{2}-s_{2}s_{3}s_{1}e^{i\delta} & c_{3}s_{1}\\
s_{2}s_{1}-c_{2}c_{1}s_{3}e^{i\delta} & -s_{1}c_{2}-s_{2}s_{3}c_{1}e^{i\delta} & c_{3}c_{1}\end{array}\right)\label{Eq. 2}\end{equation}
with natural substitutions: $\sin\theta_{23}=s_{1}$, $\sin\theta_{12}=s_{2}$,
$\sin\theta_{13}=s_{3}$, $\cos\theta_{23}=c_{1}$, $\cos\theta_{12}=c_{2}$,
$\cos\theta_{13}=c_{3}$ for the mixing angles, and $\delta$ for
the CP Dirac phase. 

Let us assume the most general symmetric mass matrix for the neutrino
sector as:

\begin{equation}
M=\left(\begin{array}{ccc}
A & D & E\\
D & B & F\\
E & F & C\end{array}\right)\label{Eq. 3}\end{equation}
and try to obtain its eigenvalues. More specifically, this reduces
to solving the set of equations:

\begin{equation}
M\mid\nu_{i}>=m_{i}\mid\nu_{i}>\label{Eq. 4}\end{equation}

\section{Neutrino mass eigenvalues}

The eigenvalues problem (4) resides in diagonalizing the matrix (3)
in order to get the eigenstates basis of the physical neutrinos, and
thus their mass eigenvalues. The procedure will lead to the following
generic solution:

\begin{equation}
m_{i}=m_{i}\left(A,B,C,\theta_{12},\theta_{23},\theta_{13}\right)\label{Eq. 5}\end{equation}
with $i=1,2,3$. In these expressions $m_{i}s$ are analytical functions
depending only on the mixing angles and the diagonal entries in the
general mass matrix. At this stage, we do not make any assumption
on the specific textures that can occur in the mass matrix when particular
symmetries are added or \emph{ad hoc} hypothesis are enforced. 

The concrete forms of $m_{i}s$ remain to be determined by solving
the following set of equations:

\begin{equation}
\begin{cases}
\begin{array}{c}
m_{1}=c_{2}^{2}A+c_{1}^{2}s_{2}^{2}B+s_{1}^{2}s_{2}^{2}C-2c_{1}c_{2}s_{2}D+2s_{1}s_{2}c_{2}E-2c_{1}s_{1}s_{2}^{2}F\\
0=c_{2}s_{2}A-c_{1}^{2}c_{2}s_{2}B-s_{1}^{2}s_{2}c_{2}C-(1-2s_{2}^{2})s_{1}E+2s_{1}s_{2}c_{1}c_{2}F\\
0=-c_{1}s_{1}s_{2}B+c_{1}s_{1}s_{2}C+c_{2}s_{1}D+c_{1}c_{2}E-(1-2s_{1}^{2})s_{2}F\\
m_{2}=s_{2}^{2}A+c_{1}^{2}c_{2}^{2}B+s_{1}^{2}c_{2}^{2}C+2c_{1}c_{2}s_{2}D-2s_{1}s_{2}c_{2}E-2c_{1}s_{1}c_{2}^{2}F\\
0=s_{1}c_{1}c_{2}B-s_{1}c_{1}c_{2}C+s_{1}s_{2}D+c_{1}s_{2}E+(1-2s_{1}^{2})c_{2}F\\
m_{3}=s_{1}^{2}B+c_{1}^{2}C+2c_{1}s_{1}F\end{array}\end{cases}\label{Eq. 6}\end{equation}

Since the actual data are not sensitive to any CP-phase violation
in the lepton sector, we have taken into account from the very beginning
$\sin^{2}\theta_{13}\simeq0$ - as it can be easily observed by inspecting
the shape of Eqs. (6) - but the proposed values for the other two
mixing angles will be embedded only in the resulting formulas for
the neutrino masses (7). 

Furthermore, one obtaines after a few manipulations the following
analytical equations:

\[
m_{1}=\frac{C\sin^{2}\theta_{12}\sin^{2}\theta_{23}-B\sin^{2}\theta_{12}\left(1+\sin^{2}\theta_{23}\right)}{\left(1-2\sin^{2}\theta_{23}\right)\left(1-2\sin^{2}\theta_{12}\right)}+\frac{A\left(1-\sin^{2}\theta_{12}\right)}{\left(1-2\sin^{2}\theta_{12}\right)},\]

\[
m_{2}=\frac{B(1-\sin^{2}\theta_{12}-\sin^{2}\theta_{23}+3\sin^{2}\theta_{12}\sin^{2}\theta_{23})-C\sin^{2}\theta_{23}\left(1-\sin^{2}\theta_{12}\right)}{\left(1-2\sin^{2}\theta_{23}\right)\left(1-2\sin^{2}\theta_{12}\right)}\]

\[
+\frac{A\sin^{2}\theta_{12}}{\left(1-2\sin^{2}\theta_{12}\right)},\]

\begin{equation}
m_{3}=\frac{C\left(1-\sin^{2}\theta_{23}\right)-B\sin^{2}\theta_{23}}{1-2\sin^{2}\theta_{23}}.\label{Eq. 7}\end{equation}

Assuming the available data concerning the mixing angles \cite{key-2}
and the mass matrix diagonal entries, one can proceed to a detailed
investigation of the resulting equations. They can reveal some interesting
features, not only with respect to the type of the mass hierarchy
(normal, inverted or degenerate) but also regarding the minimal absolute
value in the neutrino mass spectrum and the mass splitting ratio (which
imposes finally a certain relation between the diagonal entries). 

Note that some of the masses could come out negative (for certain
combinations of angles), but this is not an impediment since for any
fermion field a $\gamma_{5}\psi$ transformation can be performed
at any time, without altering the physical content of the theory.
As a result of this manipulation the mass sign changes or, equivalently,
some neutrinos have opposite CP-phases. 

Let us observe that the analytical mass equations (7) strictly impose
$\sin^{2}\theta_{12}\neq0.5$ and $\sin^{2}\theta_{12}\neq0.5$, yet
this does not forbid any closer approximation to the bi-maximal neutrino
mixing. However, in the case of solar mixing angle this behaviour
does not seem to be disturbing, since data confirm a large but not
maximal mixing. Eventually, some radiative corrections can also be
employed in order to get a more precise account for these angles,
but let us observe tha no particular mixing case is excluded \emph{ab
initio}.

These equations do not contradict the trace condition which requires
indeed a finite neutrino mass sum independently of the values of the
mixing angles. As a matter of fact, if one summs the three masses
in Eqs. (7), then the troublesome denominators disappear and the value
required by Eq. (3) is recovered. 

The particular shape of the analytical neutrino masses is due to both
the choice of the $\theta_{13}=0$ and the nonzero diagonal entries
in the mixing matrix. Any other choice - as one can observe in subsequent
section - definitely leads to a different set of equations to be solved
and, thus, to a different form of the solution.

\section{Phenomenological restrictions}

We will analyze - in the following subsections - some particular cases
of the analytical solution presented above and emphasize the most
appealing setting. We are guided in our choice by the need to obtain
plausible predictions, and even a rough estimate regarding the absolute
masses in the spectrum.

\subsection{Conserving the global lepton number $L=L_{e}-L_{\mu}-L_{\tau}$}

One of the most invoked symmetries in the lepton sector was the total
lepton number $L=L_{e}+L_{\mu}+L_{\tau}$which still holds when one
deals with Dirac neutrinos, while Majorana neutrinos violate this
symmetry with two units. Therefore, it had to be abandoned in scenarios
with Majorana neutrinos, as here is the case. 

In the particular case of conserving the global lepton number $L=L_{e}-L_{\mu}-L_{\tau}$
\cite{key-7} the shape of the mass matrix (3) becomes:\begin{equation}
M=\left(\begin{array}{ccc}
0 & D & E\\
D & 0 & 0\\
E & 0 & 0\end{array}\right)\label{Eq. 8}\end{equation}

The concrete forms of $m_{i}s$ remain to be computed by solving the
modified set of equations:

\begin{equation}
\begin{cases}
\begin{array}{c}
m_{1}=-2c_{1}c_{2}s_{2}D+2s_{1}c_{2}s_{2}E\\
0=-c_{1}s_{2}^{2}D+c_{1}s_{2}^{2}D-s_{1}c_{2}^{2}E+s_{1}s_{2}^{2}E\\
0=c_{2}s_{1}{}D+c_{2}c_{1}E{}\\
0=c_{1}c_{2}^{2}D-c_{1}s_{2}^{2}D+s_{1}s_{2}^{2}E-s_{1}c_{2}^{2}E\\
m_{2}=2c_{1}c_{2}s_{2}D-2s_{1}c_{2}s_{2}E\\
0=s_{1}s_{2}{}D+c_{1}s_{2}E{}\\
0=s_{1}c_{2}{}D+c_{1}c_{2}E{}\\
0=s_{1}s_{2}{}D+c_{1}s_{2}E{}\\
m_{3}={}0\end{array}\end{cases}\label{Eq. 9}\end{equation}
obtained straightforwardly from Eq. (6) if one puts $A=B=C=F=0$.
The lines 3, 6, 7 and 8 in Eqs. (9) express the same condition, namely:
$D=-E\cot\theta_{23}$ giving rise to a $\mu-\tau$ interchange symmetry
if $\cot\theta_{23}=-1$. The lines 2 and 4 in the set of equations
(9) are fulfiled simultaneously if and only if $\cos^{2}\theta_{12}=\sin^{2}\theta_{12}$
(maximal solar mixing angle). 

Under these circumstances, taking into consideration the maximal atmospheric
mixing angle too, the solution reads:

\begin{equation}
\left|m_{1}\right|=\left|m_{2}\right|=\sqrt{2}D\label{Eq.9}\end{equation}

\begin{equation}
m_{3}=0\label{Eq.10}\end{equation}

If the lepton number $L=L_{e}-L_{\mu}-L_{\tau}$ is rigorously conserved
the mass spectrum exhibits an inverted mass hierarchy with two degenerate
nonzero masses and bi-maximal mixing. The minimal neutrino maxx is
identical zero.

\subsection{Mass matrix with $\mu-\tau$ interchange symmetry}

Many papers \cite{key-8} develop scenarios with the $\mu-\tau$ interchange
symmetry. It seems more appealing, since the mass matrix of the neutrino
sector 

\begin{equation}
M=\left(\begin{array}{ccc}
A & D & D\\
D & B & F\\
D & F & B\end{array}\right)\label{Eq. 11}\end{equation}
can predict interesting results.

We have to insert in Eqs. (7) the restrictive condition $B=C$ and
simply express the resulting masses. They stand: \[
m_{1}=-B\frac{\sin^{2}\theta_{12}}{\left(1-2\sin^{2}\theta_{23}\right)\left(1-2\sin^{2}\theta_{12}\right)}+A\frac{\left(1-\sin^{2}\theta_{12}\right)}{\left(1-2\sin^{2}\theta_{12}\right)},\]

\begin{equation}
m_{2}=B\frac{\sin^{2}\theta_{12}}{\left(1-2\sin^{2}\theta_{23}\right)\left(1-2\sin^{2}\theta_{12}\right)}+A\frac{\sin^{2}\theta_{12}}{\left(1-2\sin^{2}\theta_{12}\right)},\label{Eq. 12}\end{equation}

\[
m_{3}=B.\]

Evidently, it is required a $\gamma^{5}$ transformation performed
on the first neutrino field in order to get the sign change for its
mass ($m_{1}$), if we assume that $A$ and $B$ have the same order
of magnitude and a suitable close-to-maximal atmospheric mixing is
invoked. In case $A\gg B$ and the atmospheric angle has a reasonable
value, no mass could need a chiral transformation to get positive
values.

The mass spectrum in the neutrino sector becomes:

\[
\left|m_{1}\right|=\left[\frac{B\sin^{2}\theta_{12}}{\left(1-2\sin^{2}\theta_{23}\right)\left(1-2\sin^{2}\theta_{12}\right)}-\frac{A\left(1-\sin^{2}\theta_{12}\right)}{\left(1-2\sin^{2}\theta_{12}\right)}\right],\]

\begin{equation}
m_{2}=\left[\frac{B\sin^{2}\theta_{12}}{\left(1-2\sin^{2}\theta_{23}\right)\left(1-2\sin^{2}\theta_{12}\right)}+\frac{A\sin^{2}\theta_{12}}{\left(1-2\sin^{2}\theta_{12}\right)}\right],\label{Eq. 13}\end{equation}

\[
m_{3}=B.\]

The physical relevant magnitudes in neutrino oscillation experiments
are the mass squared differences for solar and atmospheric neutrinos,
defined as: $\Delta m_{12}^{2}=m_{2}^{2}-m_{1}^{2}$ and $\Delta m_{23}^{2}=m_{3}^{2}-m_{2}^{2}$
respectively. They result from the above expressions (Eqs. (13)):

\begin{equation}
\Delta m_{12}^{2}\cong\frac{2AB\sin^{2}\theta_{12}}{\left(1-2\sin^{2}\theta_{12}\right)^{2}\left(1-2\sin^{2}\theta_{23}\right)}\label{Eq. 14}\end{equation}

\begin{equation}
\Delta m_{23}^{2}\cong\frac{B^{2}\sin^{4}\theta_{12}}{\left(1-2\sin^{2}\theta_{12}\right)^{2}\left(1-2\sin^{2}\theta_{23}\right)^{2}}\label{Eq. 15}\end{equation}

The mass splitting ratio defined as $r_{\Delta}=\Delta m_{12}^{2}/\Delta m_{23}^{2}$
yields in our scenario:

\begin{equation}
r_{\Delta}=2\frac{A}{B}\left(\frac{1-2\sin^{2}\theta_{23}}{\sin^{2}\theta_{12}}\right)\label{Eq. 16}\end{equation}

It is natural to presume that $A$ and $B$ have the same order of
magnitude and consequently $A/B\simeq1$. Under these circumstances
$\sin^{2}\theta_{23}\simeq0.497$ in order to fulfil the phenomenological
requirement $r_{\Delta}\simeq0.033$. 

Regarding the neutrino mass sum

\begin{equation}
\sum_{i=1}^{3}m_{i}\simeq\frac{2AB\sin^{2}\theta_{12}}{\left(1-2\sin^{2}\theta_{12}\right)\left(1-2\sin^{2}\theta_{23}\right)}\label{Eq. 17}\end{equation}
this is experimentally restricted to: $\sum_{i=1}^{3}m_{i}\sim1$eV,
if we take into consideration the Troitsk \cite{key-9} and Mainz
\cite{key-10} experiments. On the other hand, combining Eqs. (13)
and (17) one obtaines:

\begin{equation}
\sum_{i=1}^{3}m_{i}=\frac{2\sin^{2}\theta_{12}}{\left(1-2\sin^{2}\theta_{12}\right)\left(1-2\sin^{2}\theta_{23}\right)}m_{0}\label{Eq. 18}\end{equation}
with minimal neutrino mass $m_{0}=m_{3}$ . This leads to

\begin{equation}
m_{0}=\frac{\left(1-2\sin^{2}\theta_{12}\right)\left(1-2\sin^{2}\theta_{23}\right)}{2\sin^{2}\theta_{12}}\sum_{i=1}^{3}m_{i}\label{Eq. 19}\end{equation}

Assuming the phenomenological values for the sum of the neutrino masses
and the solar mixing angle $\sin^{2}\theta_{12}\simeq0.31$ one can
analyze the behaviour of the $m_{0}$ in terms of the atmospheric
mixing angle by studying the function:

\begin{equation}
m_{0}(\sin^{2}\theta_{23})=0.613\left(1-2\sin^{2}\theta_{23}\right)\sum_{i=1}^{3}m_{i}\label{Eq. 20}\end{equation}

A plausible value (with the above considered values for mixing angles)
can now be inferred: $m_{0}\simeq0.0035$eV. It is very close to the
value obtained by the author (second reference in \cite{key-4}) in
a particular 3-3-1 model where the diagonal entries of the neutrino
mass matrix were obtained in a specific manner, without resorting
to any additional symmetry.

\section{Concluding remarks}

In this paper we have proved that the exact mass-eigenstates of a
general neutrino mass matrix with no CP-phase violation can be exactly
computed. The results accomodate the observed solar mixing angle and
exclude the exact maximal mixing for the atmospheric angle, but do
not forbid any closer approximation for such a setting. Therefore,
they could be in good agreement with the data and can predict the
correct mass splitting ratio. Our predictions also include the inverted
mass hierarchy in the neutrino sector and the minimal absolute mass
- $m_{0}\simeq0.0035$ eV - since the $\mu-\tau$ interchange symmetry
is employed. The global lepton symmetry $L=L_{e}-L_{\mu}-L_{\tau}$supplies
two degenerate nonzero masses and one identical to zero, within the
inverted hierarchy as well. However, the general case can be ragarded
as a perturbation that softly breaks this lepton symmetry by introducing
small nonzero diagonal entries. The amazing feature seems to be the
unexpected similarity of these general results with the ones obtained
by the author in a particular 3-3-1 model with specific diagonal entries
(proportional to charged lepton masses).

\end{document}